# Seismicity prior to the 2016 Kumamoto earthquakes

Author #1: K. Z. Nanjo, Global Center for Asian and Regional Research, University of Shizuoka, 3-6-1, Takajo, Aoi, Shizuoka 420-0839, Japan, Tel: +81-54-245-5600, Fax: +81-54-245-5603, E-mail: nanjo@u-shizuoka-ken.ac.jp

Author #2: J. Izutsu, International Digital Earth Applied Science Research Center, Chubu Institute for Advanced Studies, Chubu University, 1200 Matsumoto-cho, Kasugai, Aichi 487-8501, Japan, Tel: +81-568-51-1111, Fax: +81-568-51-1642, E-mail: izutsu@isc.chubu.ac.jp

Author #3: Y. Orihara, Department of Physics, Tokyo Gakugei University, 4-1-1, Nukuikita, Koganei, Tokyo 184-8501, Japan, Tel +81-42-329-7484, Fax: +81-42-329-7484, E-mail: orihara@u-gakugei.ac.jp

Author #4: N. Furuse, Institute of Oceanic Research & Development, Tokai University, 3-20-1, Orido, Shimizu, Shizuoka 424-0902, Japan, Tel: +81-54-337-0946, Fax: +81-54-336-0920, E-mail: furuse@sems-tokaiuniv.jp

Author #5: S. Togo, Department of Physics, Tokyo Gakugei University, 4-1-1, Nukuikita, Koganei, Tokyo 184-8501, Japan, Tel +81-42-329-7484, Fax: +81-42-329-7484, E-mail: f117126w@st.u-gakugei.ac.jp

Author #6: H. Nitta, Department of Physics, Tokyo Gakugei University, 4-1-1, Nukuikita, Koganei, Tokyo 184-8501, Japan, Tel +81-42-329-7484, Fax: +81-42-329-7484, E-mail: a120349w@st.u-gakugei.ac.jp

Author #7: T. Okada, Department of Physics, Tokyo Gakugei University, 4-1-1, Nukuikita, Koganei, Tokyo 184-8501, Japan, Tel +81-42-329-7484, Fax: +81-42-329-7484, E-mail: b142312x@st.u-gakugei.ac.jp

Author #8: R. Tanaka, Department of Physics, Tokyo Gakugei University, 4-1-1, Nukuikita, Koganei, Tokyo 184-8501, Japan, Tel +81-42-329-7484, Fax: +81-42-329-7484, E-mail: f117124x@st.u-gakugei.ac.jp

Author #9: M. Kamogawa, Department of Physics, Tokyo Gakugei University, 4-1-1, Nukuikita, Koganei, Tokyo 184-8501, Japan, Tel +81-42-329-7484, Fax: +81-42-329-7484, E-mail: kamogawa@u-gakugei.ac.jp

Author #10: T. Nagao, Institute of Oceanic Research & Development, Tokai University, 3-20-1, Orido, Shimizu, Shizuoka 424-0902, Japan, Tel: +81-54-337-0946, Fax: +81-54-336-0920, E-mail: nagao@scc.u-tokai.ac.jp

Corresponding author: T. Nagao

**Abstract**

We report precursory seismic patterns prior to the 2016 Kumamoto earthquakes, as measured by four different methods based on changes in seismicity that can be used for earthquake forecasting: the $b$-value method, two methods of seismic quiescence evaluation, and an analysis of seismicity density in space and time. The spatial extent of precursory patterns differs from one method to the other and ranges from local scales (typically, asperity size) to regional scales (e.g., 2° × 3° around the source zone). The earthquakes were preceded by periods of pronounced anomalies, which lasted in yearly scales (1.5 years), or longer (> 3 years). We demonstrate that a combination of multiple methods detected different signals prior to the Kumamoto earthquakes. This indicates great potential to reduce the hazard at possible future sites of earthquakes relative to long-term seismic hazard assessment. We also found that the seismic quiescence in a regional scale area, detected by using the two methods of seismic quiescence evaluation, was a common precursor to the 2016 Kumamoto earthquakes and 2015 Off Satsuma Peninsula earthquake. The result allows us to interpret both events as the onset that occurred at a section along the tectonic line from the Okinawa Trough through the Beppu-Shimabara graben.

**Introduction**

The 2016 Kumamoto earthquakes, including a magnitude $M$7.3 mainshock that struck on April 16, 2016, as well as active foreshocks and aftershocks, fulfilled a section of the Futagawa-Hinagu fault zones. The zones are encompassed by the Beppu-Shimabara graben, a geological formation that runs across the middle of Kyushu, from Beppu Bay in the east to the Shimabara Peninsula in the west. This is understood as being the result of crustal deformation caused by the rifting and spreading of the Okinawa Trough, which is viewed as a continuation of the Beppu-Shimabara graben (Tada, 1985). The November 14, 2015 $M$7.1 earthquake occurred off Satsuma Peninsula in the Okinawa Trough. This event and the Kumamoto earthquakes are considered to be the onset that occurred under the same tectonics (Ishibashi, 2016).

A wide variety of approaches have been applied to earthquake prediction and forecasting.

Most proposed prediction and forecasting methods rely on the concept of a diagnostic precursor; i.e., some kind of signal observable before earthquakes that indicates with high probability the location, time, and magnitude of an impending event. Precursor methods include changes in strain rates, seismic wave speeds, and electrical conductivity; variations of radon concentrations in groundwater, soil, and air; fluctuations in groundwater levels; electromagnetic variations near and above the Earth's surface; and seismicity patterns. Despite this, the search for diagnostic precursors has not yet produced a successful short-term prediction scheme (e.g., Keilis-Borok, 2002; Scholz, 2002; Kanamori, 2003; International Commission on Earthquake Forecasting for Civil Protection, 2011).

Despite such a notable lack of success, there has been a resurgence of research on earthquake predictability motivated by better monitoring networks and data on past events, new knowledge of the physics of earthquake ruptures, and a more comprehensive understanding of stress evolution and transfer. The California and international working groups, RELM (Field 2007) and CSEP (Jordan, 2006), have been supporting research on earthquake predictability, conducting scientific experiments under rigorous, controlled conditions and evaluating them using accepted criteria specified in advance. These groups point to future directions of model development of earthquake prediction and forecasting as well as its testing. However, making full use of currently available resources for and a new knowledge and understanding of seismology, there needs predictability research that is not directly associated with the RELM and CSEP, but rather that addresses more general questions about the improvement of different hypotheses on earthquake generation and different concepts related to diagnostic precursor.

Different methods exist to measure, map, and evaluate possible episodes before mainshocks. Examples include the $b$-value method (e.g., Schorlemmer and Wiemer, 2005; Nanjo et al., 2012), RTL/RTM-algorithms (e.g., Sobolev and Tyupkin, 1997; Nagao et al., 2011), the *Z*-value method (e.g., Wiemer and Wyss, 1994; Wyss et al., 2004; Katsumata, 2015), and seismicity density analysis (Lippiello et al., 2012). To our knowledge, there is no standardized approach that encourages researchers to rely exclusively on a single method. Instead, multiple methods must be used to find evidence of precursory episodes. One disadvantage is that the results obtained by different methods may not be easily compared, although one may gain more confidence in impending earthquakes when using different methods. Also, additional insight may be gained because of intrinsic differences in the statistical characterization of seismic patterns. Early attempts were made by Wyss et al. (2004) who used the RTL-algorithm and *Z*-value methods and by Enescu and Ito (2001) who used the *b* value and *Z*-value methods, and a fractal dimension approach.

This paper reports the first results of recognizing seismic patterns as possible precursory episodes to the 2016 Kumamoto earthquakes using existing four different methods: the *b* value method, two kinds of seismic quiescence evaluation methods (RTM-algorithm and *Z*-value), and seismicity density analysis. These methods are based on different assumptions, selection of sampling volumes, algorithms, and definitions of anomalies, thus they are representative of a wide variety of methods that can be used to detect precursory anomalies. Before presenting further details of our study, we present a brief overview of the methods that were used.

**Methods**

*The b value method*: The *b* value is the slope of the Gutenberg-Richter (GR) frequency-magnitude distribution of earthquakes (Gutenberg and Richter, 1944), $\log_{10}N = a - bM$, where $N$ is the cumulative number of earthquakes of $M$ or greater, $a$ is the earthquake productivity of a volume, and *b* is used to describe the relative occurrence of large and small events (i.e., a high *b* value indicates a larger proportion of small earthquakes, and vice versa). Spatial and temporal changes in *b* are known to reflect a structural heterogeneous structure (e.g., Mogi, 1962) such as strong coupling along subduction zones and a magma chamber (e.g., Tormann et al., 2015). In the laboratory and the Earth's crust, the *b* value is also known to be inversely dependent on differential stresses (Scholz, 1968, 2015). In this context, measurements of spatial temporal changes in *b* could act as a stress meter to help image asperities, the highly stressed patches in faults where future ruptures are likely to occur (e.g., Schorlemmer and Wiemer, 2005; Nanjo et al., 2012; Tormann et al., 2015).

*The RTM-algorithm*: This is a type of weighted method that assesses the time, space and the size of earthquakes, and is a modified RTL algorithm (e.g., Sobolev and Tyupkin, 1997, 1999, Huang and Nagao, 2002, Huang, 2004, 2006, Sobolev, 2011). The RTM-algorithm measures the level of seismic activity in moving windows by counting the number of earthquakes that are weighted by their size and inversely weighted by their distances in time and space from the point of observation. A detailed description is provided by Nagao et al. (2011). A decrease of the *RTM* value implies a decrease in seismicity compared to the background rate around the investigated location (a seismic quiescence). A recovery stage from quiescence to the background level can be broadly considered as foreshock activation. The RTM-method evaluates both seismic quiescence and the following stage of activation. In addition, the location of extreme, if it is an anomaly, can be found by performing RTM-calculations with the centers of the sampling circles at the nodes of a grid.

*The Z-value method*: The *Z*-mapping approach measures the difference in seismicity rate, within moving time windows, to the background rate by a standard deviation, *Z* (Wiemer and Wyss, 1994; Wyss et al., 2004; Katsumata, 2015). The purpose is to detect possible periods of anomalously low seismicity just before the mainshock near its epicenter, and to evaluate the statistical significance of such quiescence compared with all other parameters that may have occurred at random times and locations. The standard deviation $Z$ is defined as $Z = (R_1 - R_2)/(S_1/n_1 + S_2/n_2)^{1/2}$, where $R_1$ and $R_2$ are the mean rates, $S_1$ and $S_2$ are the variances, and $n_1$ and $n_2$ are the number of events in the first and second periods to be compared. The larger the $Z$-value, the more significant the observed difference.

*Seismicity density analysis*: An increase in the number of smaller magnitude events before large earthquakes is often observed. The linear density probability of earthquakes occurring before and after events defined as mainshocks, $\rho(\Delta r)$, as a function of the distance from the mainshock hypocenter, $\Delta r$, displays a symmetrical behavior (Lippiello et al., 2012). This behavior indicates that the size of the area fractured during the mainshock is associated with the spatial organization of seismicity before the mainshock, because the spatial organization after the mainshock reflects the mainchock rupture area (Lippiello et al., 2012). The analysis based on Lippiello et al. (2012), hereinafter referred to as the seismicity density analysis, can be implemented to forecast the size of the mainshock. The objective of this paper is to detect such symmetric behavior for the Kumamoto earthquakes and to evaluate the possible size of the preparatory areas to the earthquakes.

**Data**

Our dataset is the earthquake catalog maintained by the Japan Meteorological Agency (JMA). We mainly used earthquakes since 2000, the start of modernization of the seismic networks feeding data to JMA (Obara et al., 2005). Since aftershocks add noise to both the *Z*-value method and the RTM-algorithm, we used a declustered catalog as a basic input for these methods. We eliminated aftershocks using the JMA standard declustering program (For details, see Appendix 1 and Fig. A1 in Additional file 1), which is classified as a link algorithm (Frolich and Davis, 1990).

A reliable estimate of completeness magnitude $M_c$, above which all earthquakes are considered to be detected by a seismic network, is vital for seismicity-related studies. We paid attention to catalog completeness, and performed a completeness check as a pre-processing step of individual analyses while referring to a comprehensive analysis of $M_c$ in Japan (Nanjo et al., 2010).

**Results**

The *b* value method

A map view (Fig. 1a) of *b* values based on seismicity over a period from 2000 to April 14, 2016 21:25 (immediately before the Kumamoto earthquakes) reveals a zone of low *b* values near the eventual epicenters of the mainshock and two *M*6.5-class foreshocks. The characteristic dimension of unusually low *b* values in an approximately 0.3° × 0.3° area in Fig. 1a is ≈ 30 km. The *M*5 earthquake on June 8, 2000 occurred near the mainshock hypocenter so that we excluded seismicity related to this event and created maps of *b* values (Fig. A2 in Additional file 1). These maps show that a zone of low *b* values is insensitive to this exclusion, indicating that the 2000 *M*5 earthquake is not the main cause of the low spatially-mapped *b* values in Fig. 1. Comparison with the mainshock rupture zone (National Research Institute for Earth Science and Disaster Resilience, 2016) shows the influence of structural heterogeneity on spatial distribution in *b* values in such a way that rupture propagation of the mainshock terminated an area near Mt. Aso with *b* ~ 1, higher than *b* values within the rupture zone ($b$ = 0.6~0.8). The high-temperature area around the magma chamber of Mt. Aso may have contributed to termination of the rupture (Yagi et al., 2016) and high $b$ values (e.g., Roberts et al., 2015). We compared the map in Fig. 1a with the sequence of foreshocks, mainshock, and aftershocks with $M$5 or greater. Areas of low $b$ values contained 84% of the sequence (Fig. A3 in Additional file 1). Two of the three aftershock regions, which lay further to the northeast, were observed around the high $b$ value zones near Mt. Aso and Mt. Yufu, as observed in previous studies about $b$ value characterization of volcanic systems (e.g., Roberts et al., 2015).

For earthquakes in the cylindrical volume with a radius $R$ = 10 km, centered at the epicenter of the mainshock (Fig. 1c), $b$ values after the fluctuation due to aftershocks of the 2000 $M$5 earthquake were stable (See also Fig. A4 in Additional file 1). The precursory duration of the $b$-value is unknown, because the time variation in $b$ shows no clear change-point during the analyzed period. A similar time-series analysis was performed for seismicity in the Futagawa, Hinagu, Aso-Kuju, and Beppu fault zones (See Figs. A5-A9 in Additional file 1). The $b$ values show no systematic increase or decrease for all fault zones. A summary of the results is provided in Table 1.

RTM-algorithm

As described in the Data section, we first eliminated aftershocks. Next, we conducted a completeness analysis of the JMA catalog for the study region (Fig. 2) and obtained typically $M_c$ = 2 so that we primarily set a lower threshold at $M_{min}$ = 2. Then, through an extensive parameter survey, we found anomalous seismic quiescence patterns as explained next.

A temporal variation of *RTM* at an almost equidistant point from the 2015 Off Satsuma Peninsula earthquake and the 2016 Kumamoto mainshock showed a significant decrease prior to the occurrence of both (Fig. 2a, see also Table A1 in Additional file 1). A deviation from the background level started at 2014.8 years for mid-October 2014 and the strongest deviation in 2015.8 years for mid-October 2015 was about -20. During the critical period, the $R$, $T$, and $M$ functions attained values of about -2.5~-3.0. During a recovery stage from the quiescence to the background level, the 2015 Off Satsuma Peninsula earthquake and the 2016 Kumamoto mainshock occurred. This property is similar to that documented by other studies that used the RTL/RTM-algorithm (e.g., Sobolev and Tyupkin, 1997, 1999, Huang and Nagao, 2002, Huang, 2004, 2006; Nagao et al., 2011, Sobolev, 2011). Since the RTM algorithm is statistical and nonlinear, we were unable to identify which of earthquakes contributed to the recovery stage; this topic is beyond the scope of our work in this paper.

A map view of $RTM$ values on October 1, 2015 reveals a highly significant change between the 2015 Off Satsuma Peninsula earthquake and the 2016 $M$7.3 Kumamoto mainshock indicated by stars in Fig. 2b, in which the parameter set was the same as that for Fig. 2a. The characteristic dimension of seismic quiescence in an approximately 2º × 3º area is 200~300 km. We created a map of $RTM$ values using the same parameter set as that for Fig. 2b except that a higher threshold ($M_{min}$ = 2.4) was applied. The position of the anomalous zone is not very sensitive (See Panel E of Fig. A10 in Additional file 1). We also used three different sets of parameters with $M_{min}$ = 2 and created $RTM$ maps to support a more robust anomalous pattern (See Panels A-C of Fig. A10 and Table A1 in Additional file 1).

To strengthen the related conclusion, it would be better to conduct a reliability analysis of the revealed anomalies in detail, e.g., following the approach taken by Huang (2006). However, the main purpose of this paper was to provide the first results on how to recognize possible precursory episodes to the 2016 Kumamoto earthquakes, as described in the Introduction. Thus, we presented essential parts of the reliability analysis such as completeness check, quality test of a declustered catalog, and some levels of a parameter survey. For a full justification of the present conclusion, further detailed investigations involving the robustness of the temporal and spatial pattern should be conducted.

Our results do not indicate that the observed seismic quiescence precedes either of the mainshocks. The 2015 Off Satsuma Peninsula earthquake and the 2016 Kumamoto earthquakes are considered to be the onset that occurred at a section along the tectonic line from the Okinawa Trough through the Beppu-Shimabara graben (Ishibashi, 2016). Our results, combined with these tectonics, indicate that a series of earthquakes along the tectonic line was preceded by seismic quiescence. These results are summarized in Table 1.

The Z-value method

Similar to the RTM-algorithm, the processing steps were performed to ensure data quality. An extensive parameter survey was then conducted to choose the most suitable parameter set. The cumulative number of earthquakes with $M_{min} = 2.4$ as a function of time for a circle with sampling radius $R = 100$ km, centered at a point about 60 km away to the northeast from the epicenter of the 2015 Off Satsuma Peninsula mainshock, showed an anomaly of a few earthquakes during about 1 year before this mainshock (Fig. 3) and about 1.5 years before the Kumamoto mainshock. With a look-ahead time window $T_w = 1$ year, sampling radius $R = 100$ km, and $M_{min} = 2.4$, comparison of the seismicity rate during the last year before the Off Satsuma Peninsula mainshock with the background seismicity rate resulted in $Z = 2.9$. This representative result shows a typical precursory quiescence pattern to a large earthquake. At other points, typical precursory pattern showing fluctuation in $Z$ at low levels followed by a rapid increase to high values is degraded or unobservable, as shown in Fig. A11 in Additional file 1 (see a further parameter survey in Appendix 2 in Additional file 1 that supports our results). Fig. 3 also shows that during the fluctuation in $Z$, $Z$ is below zero at most times, indicating preceding seismic activation to quiescence. Following the RTM-algorithm, we used a lower threshold at $M_{min}$ = 2.0. Quiescence patterns were observed but this characteristic was degraded (Fig. A12 in Additional file 1). In summary, in the area with $R$ = 100 km centered at a point about 60 km away to the northeast from the epicenter of the Off Satsuma Peninsula earthquake, the $Z$-value analysis detected a typical precursory quiescence, which lasted about 1 year (Table 1).

Seismicity density analysis

Before assuming the approach based on Lippiello et al. (2012) for Japan, a complete check of the JMA catalog since 1998 was performed as a pre-processing step. We decided to use inland earthquakes only and set a lower threshold at $M_{min}$ = 1. The previous $Z$-value and RTM approaches were applied to data including both inland and offshore earthquakes (Figs. 2 and 3). $M_c$ of the JMA catalog was generally lower in inland regions than offshore regions (Nanjo et al., 2010), which explains why $M_{min}$ = 1 for this seismicity density analysis is different from $M_{min}$ = 2 and 2.4 for the $Z$-value and RTM approaches.

As shown in the inset of Fig. 4 for mainshock magnitudes $4 \leq m < 5$, $\rho(\Delta r)$ before and after mainshocks was very similar in the whole spatial range for all magnitude ranges $3 \leq m < 6$, consistent with Lippiello et al. (2012). The decay of seismicity density with distance is well modeled by an inverse power-law $\rho(\Delta r) \sim \Delta r^{-\eta}$ at $\Delta r \gg 0$, where $\eta$ is a constant of 1~2 (Richards-Dinger et al., 2010). We assumed a typical $\eta = 1.35$ (Felzer and Brodsky, 2006).

We next compared the size of spatial organization before and after mainshocks with the size of asperity (Fig. 4). As a representative of the former size, we detected the deviation point from a scaling relation $\rho(\Delta r) \sim \Delta r^{-\eta}$ (the inset in Fig. 4). This deviation point is defined as the characteristic distance $\Delta r_c$, below which the scaling is no longer valid, due to large variance of seismicity density or low seismicity density at very short distances to the mainshock hypocenter, above a location uncertainty of ~100 m in the JMA catalog. We created a plot of $\Delta r_c$ versus $m$ for $3 \leq m < 6$ in Fig. 4. The latter size was based on the scaling between asperity area $S_a$ and $m$ (Somerville et al., 2015). Assuming a circular asperity $S_a = \pi l_a^2$ where $l_a$ is the characteristic asperity radius, we included the $l_a$-$m$ relation in Fig. 4. Data (grey points) show a positive correlation with large scatter, and the $\Delta r_c$-$m$ correlation appears to be similar to the $l_a$-$m$ relation. To support this idea, data from the $\Delta r_c$-$m$ correlation needs to be increased.

We then considered the Kumamoto case in Fig. 5. When using the same approach as Lippiello et al. (2012), the $M$7.3 shock and a preceding $M$6.5 shock were close to each other in space and time of the seismicity density analysis for $m$=6~7. Therefore, we used seismicity subsequent to the $M$7.3 quake and seismicity prior to the $M$6.5 quake. Although Fig. 5 has a slightly large scatter, data (blue stars) of seismicity subsequent to the $M$7.3 quake are approximated by $\rho(\Delta r) \sim \Delta r^{-\eta}$ with $\eta = 1.35$ for distances $\Delta r \geq 8$ km ($\Delta r_c = 8$ km). As a reference, we applied the same procedure to seismicity subsequent to the $M$6.5 shock until $M$7.3 shock (red triangles) and $\rho(\Delta r) \sim \Delta r^{-\eta}$ with $\eta = 1.35$ being a reasonable approximation for $\Delta r \geq 4$ km ($\Delta r_c = 4$ km). Analysis of seismicity 3, 6, 12, and 18 years before the $M$6.5 shock (inset of Fig. 5) reveals that the inverse power-law is an approximation for data at distances around $\Delta r \geq 40$ km ($\Delta r_c = 40$ km), with density levels above the background implied from distant seismicity (grey shading). An inverse power-law with $\eta = 1.35$ was clearly observed for the 3 year period, although the exponent $\eta$ showed a gradual increase over the past 18 years. Data of $\Delta r_c$ for the Kumamoto case are included in Fig. 4. Although this figure still has a slightly large scatter, it is reasonable that a positive correlation between $\Delta r_c$ and $m$ is comparable with the $l_a$-$m$ relation. The similar dependences on $m$ indicate that the size of the spatial organization of seismicity is governed by the size of asperity (see Table 1 for a summary).

**Discussion and conclusions**

Although long-term probabilistic seismic-hazard assessment can be made in Japan (e.g., National Seismic Hazard Maps for Japan) and other seismically active regions, it is generally accepted that immediate local precursory phenomena are not seen ubiquitously. In order to test the hypothesis of a local precursor for a fault system, the Parkfield Earthquake Predictability Experiment was initiated in 1985. The expected Parkfield earthquake occurred beneath the heavily instrumented region on September 28, 2004. No local precursory changes were observed (Lindh, 2005). In the absence of local precursory signals, the next question is whether broader anomalies develop and in particular whether there is anomalous seismicity.

Before attempting to identify precursory phenomena, a fundamental question is whether there are different forecast methods applicable to seismicity in a region of interest, in particular in Japan. This is the question that we addressed in this paper. We decided to use the *b*-value method, the RTM-algorithm, seismicity density analysis, and the *Z*-value method, and applied them to seismicity before the Kumamoto earthquakes. Our study is the first report on this theme after completing the modernization of seismic monitoring in Japan. Before this, an early attempt was made by Enescu and Ito (2001), who used the *b*-value method, the *Z*-value method, and fractal dimensions to discover premonitory quiescence followed by activation of seismicity about 2 years before the occurrence of the 1995 *M*7.2 Kobe earthquake. These authors conducted their study before the completion of the modernization of seismic monitoring in Japan and used the local network data of Kyoto University, besides the JMA catalog data, to ensure a relatively low and stable magnitude of completeness during the whole analyzed time period.

The combination of multiple methods may enhance the reliability of the revealed anomalies when compared to the reliability of a single method. We performed some comprehensive comparisons of the results from each method to support this claim. The properties of precursory episodes are summarized in Table 1. Estimates of the duration and spatial extent contain uncertainties, which depend on the approaches taken in the analysis. The durations were almost identical in both the RTM-algorithm and the *Z*-value method. In these methods, the resulting maximum expression of the anomalies did not coincide with the epicenters of the Kumamoto mainshock, nor of the Off Satsuma Peninsula mainshock. Furthermore, both methods did not identify the same locations for the anomalous maximum. Wyss et al. (2004) indicated a reason for the difference between methods in pinpointing the location of an anomaly. The weighting of results based on the size of the earthquakes, which is only done in the RTM method but not in the *Z*-value method, results in some differences in the estimated significance in most samples. Therefore, the maximum expression of the anomaly was not observed at exactly the same locations. The reason for the maximum expression of anomaly not pointing to the epicenters may be due to the fact that the onset of both earthquakes may have been at a segment along the tectonic line from the Okinawa

Trough (Ishibashi, 2016). The large area of the anomaly may reflect the nature of the process leading up to the phenomenon in the regional tectonic condition. Nonetheless, the two methods based on different assumptions, different algorithms, and different definitions of anomaly, arrived at very similar results. This strongly suggests that the observed anomalies are real, and can be determined with considerable reliability.

In contrast to these two methods, the *b*-value method and seismicity density analysis seem to have narrower spatial extents (Table 1). The close match found in spatial extent between the different approaches implies that asperities (highly stressed patches) may act as an indicator of the preparation process to an impending earthquake. A duration of 3 years or longer from the seismicity density analysis (Table 1) is longer than that revealed by both the RTM-algorithm and the *Z*-value method. The precursory duration from application of the *b*-value method to analyzed data for the past 16 years is unknown. The timing of the earthquake remains uncertain from low *b* values and spatial organization prior to the Kumamoto earthquakes. This is consistent with another observation of active faults in which the heterogeneous pattern of *b* values at Parkfield was, to a high degree, stationary for the past 35 years and the 2004 *M*6 earthquake eventually occurred at a zone of low *b* values (Schorlemmer and Wiemer, 2005). A decrease in *b* tracking stress buildup, as expected from a laboratory experiment (Scholz, 1968), may not be observable for the decade-scale observation of active faults in Japan and California, because it is too short to observe such a decrease in *b*. Instead, the stationary nature of *b* values, as observed in Fig. 1, is reasonable. The mechanism of stress buildup within the fault zones is uncertain, but one hypothesis is that a steady slip of the deeper continuation of faults that does not produce earthquakes, but still involves motions across the fault, forces the upper crust around the faults to deform and thus concentrate stress. This is still difficult to measure directly.

Overall, our findings indicate that the methods we adopted do not allow the Kumamoto earthquakes to be predicted exactly. The spatial extent of precursory patterns that were detected differed from one method to another and ranged from local scales (equivalent to an asperity dimension) to regional scales (typically, 100~200 km). The Kumamoto earthquakes were preceded by periods of pronounced anomalies, which lasted yearly scales (1.5 years) or longer (> 3 years). Given the widely different scales of anomalies in time and space, a combination of multiple methods was able to detect different signals prior to the Kumamoto earthquakes. This strongly suggests a great potential to reduce the hazard at possible future sites of earthquakes relative to long-term seismic hazard assessment. We also found that the seismic quiescence in a regional scale area, detected by using the RTM-algorithm and the *Z*-value method, was a common precursor to the 2016 Kumamoto earthquakes and 2015 Off Satsuma Peninsula earthquake that occurred five months before the 2016 Kumamoto earthquakes. The precursory durations of 1 year to the 2015 Off Satsuma Peninsula earthquake were almost identical in both approaches. The result allows us to interpret both events as the onset that occurred at a section along the tectonic line from the Okinawa Trough through the Beppu-Shimabara graben (Ishibashi, 2016).

**Competing interests**

The authors declare no competing financial interests.

**Authors' contributions**

All the authors performed numerical simulations, analyzed data and prepare the figures:

in particular, *b* value method (KZN, JI), RTM-algorithm (TN), *Z*-value method (KZN), seismicity density analysis (MO, ST, HN, TO, RT, MK), and application of declustering programs (KZN, NF, TN). KZN and TN helped to draft the manuscript. KZN wrote the final manuscript. All the authors read and approved the final manuscript.

**Acknowledgements**

The authors thank the Editor (M. Hashimoto) and two anonymous referees for their useful comments. This study was partly supported by the Ministry of Education, Culture, Sports, Science and Technology (MEXT) of Japan, under its Earthquake and Volcano Hazards Observation and Research Program (KZN, JI, MO, MK, TN) and a Grant-in-Aid for Scientific Research (C), No. 26350483, 2014-2017 (MK), and the Collaboration Research Program of IDEAS, Chubu University IDEAS201614 (JI).

**Table 1.** Characteristic precursory phenomena

| Method | Spatial extent | Duration |
|---|---|---|
| $b$-value method | Area of 0.3° × 0.3° containing a part of the source zone of the Kumamoto mainshock | Unknown |
| RTM-algorithm | Area of 2° × 3° between the Kumamoto and Off Satsuma Peninsula mainshocks | 1.5 years for the Kumamoto mainshock (1 year for the Off Satsuma Peninsula mainshock) |
| $Z$-value method | Area with $R = 100$ km between the Kumamoto and Off Satsuma Peninsula mainshocks | 1.5 years for the Kumamoto mainshock (1 year for the Off Satsuma Peninsula mainshock) |
| Seismicity density analysis | Area with $\Delta r = 8{\sim}40$ km containing a part of the source zone of the Kumamoto mainshock | 3 years or longer for the Kumamoto mainshock |

**Figure legends**

**Fig. 1**. **a** Map of *b* values obtained from events from January 1, 2000 to April 14, 2016 21:25 (immediately before the Kumamoto earthquakes). Overlapped on the map are the hypocenter (yellow star) and rupture zone (rectangle) of the *M*7.3 mainshock; *M*6.5-class foreshocks (red stars); $M \geq 5$ foreshocks (squares); and $M \geq 5$ aftershocks (circles). Triangles: volcanos; red line: faults. We calculated *b* values (Woessner and Wiemer, 2005) for events with depths 25 km or shallower falling in a cylindrical volume with radius $R = 5$ km, centered at each node on a 0.02º × 0.02º grid (we did not use a fixed number of earthquakes) and plotted a *b* value at the corresponding node only if at least 50 events in the cylinder yielded a good fit to the GR law. **b** GR fitting is shown for points A, B, and C in **a**. **c** Plot of *b* values as a function of time, as obtained from seismicity in the cylindrical volume with $R = 10$ km, centered at the mainshock. We used a moving window approach, whereby the window covered 500 events. Uncertainty estimates are according to Shi and Bolt (1982). Grey squares and open circles show the *b* values obtained from foreshocks and aftershocks, respectively. Horizontal line: regional average $b = 0.93$; vertical line: moment of the mainshock.

**Fig. 2**. **a** Temporal variations of the *RTM* (black), *R* (blue), *T* (green), and *M* (red) at the point indicated by a cross in **b**. Red arrows show the moments of the 2015 Off Satsuma Peninsula earthquake and the mainshock of the 2016 Kumamoto earthquakes; both epicenters are indicated by stars in **b**. **b** Areas of seismic quiescence before the two earthquakes (stars). The figure, generated by a grid with 0.1º × 0.1º spacing, shows the most quiescent period on October 1, 2015, indicated by a black arrow in **a** (45 days before the 2015 Off Satsuma Peninsula earthquake). The cross is the point of time variation shown in **a**. Parameters used were $R_0 = 100$ km, $R_{max} = 200$ km, $T_0 = 1$ year, $T_{max} =$

2 years, and $M_{min}$ = 2 (Table A1 in Additional file 1). For parameter definition, see Nagao et al. (2011).

**Fig. 3**. Cumulative number of earthquakes (blue, scale on left) with $M \geq 2.5$ as a function of time, up to the occurrence of the Off Satsuma Peninsula earthquake in 2015.8 (star). Curve in red shows the Z-values (scale on right), resulting from every position of the time window $T_w$ = 1 year. The sample is from a circle with $R$ = 100 km, centered at the black dot in the inset. Also included in the inset for reference are stars showing the 2015 Off Satsuma Peninsula and the 2016 Kumamoto mainshocks, and grey dots showing the centers of the collecting volumes with $R$ = 100 km to create Fig. A11 in Additional file 1.

**Fig. 4.** Plot of the characteristic distance $\Delta r_c$ as a function of mainshock magnitude $m$ for seismicity prior (circles) and subsequent (diamonds) to mainshocks in Japan. The Kumamoto earthquakes (see also Fig. 5) are shown as a set of data of prior and subsequent seismicity as follows: blue line segment from $m$6.5 to 7.3 with a blue circle, indicating seismicity during a 3-year period prior to the $M$6.5 shock; blue line segment from $m$6.5 to 7.3 with a blue diamond, indicating seismicity subsequent to the $M$7.3 shock. As a reference, data for seismicity in the period subsequent to the $M$6.5 shock until the $M$7.3 shock is included (red diamond). Thin solid line is drawn by using extrapolation from the scaling (thick solid line) of the characteristic asperity radius $l_a$ with $m$ of 6.6 to 9.2, based on Somerville et al. (2015). Thin dashed lines are drawn by using extrapolation from the one-standard-deviation limits (thick dashed line). The inset shows a plot of $\rho(\Delta r)$ as a function of $\Delta r$ from the mainshock with $m$ = 4~5 for seismicity prior (circles) and subsequent (diamonds) to mainshocks. Data are fitted with $\rho(\Delta r) \sim \Delta r^{-\eta}$ with $\eta = 1.35$. The deviation point, below which an inverse power-law is no longer valid, is marked by $\Delta r_c$.

**Fig. 5.** Plot of $\rho(\Delta r)$ as a function of $\Delta r$ for the Kumamoto earthquakes. Blue stars: seismicity subsequent to the *M*7.3 shock. As a reference, data on seismicity subsequent to the *M*6.5 shock until the *M*7.3 shock are marked by red upward-pointing triangles. Also included in this figure is an inverse power law with an exponent $\eta = 1.35$. The characteristic distance $\Delta r_c$, below which an inverse power-law is no longer valid, is marked for respective data. The inset shows a plot for seismicity prior to the *M*6.5 shock for different periods: 18 years (crosses), 12 years (grey downward-pointing triangles), 6 years (yellow diamonds), and 3 years (blue circles). Data for the 3 year period are fitted with an inverse power-law with $\eta = 1.35$. The deviation point, below which an inverse power-law is no longer valid, is marked by $\Delta r_c$. Grey shading indicates distances larger than 55 km, beyond which data are referred to as background density levels.

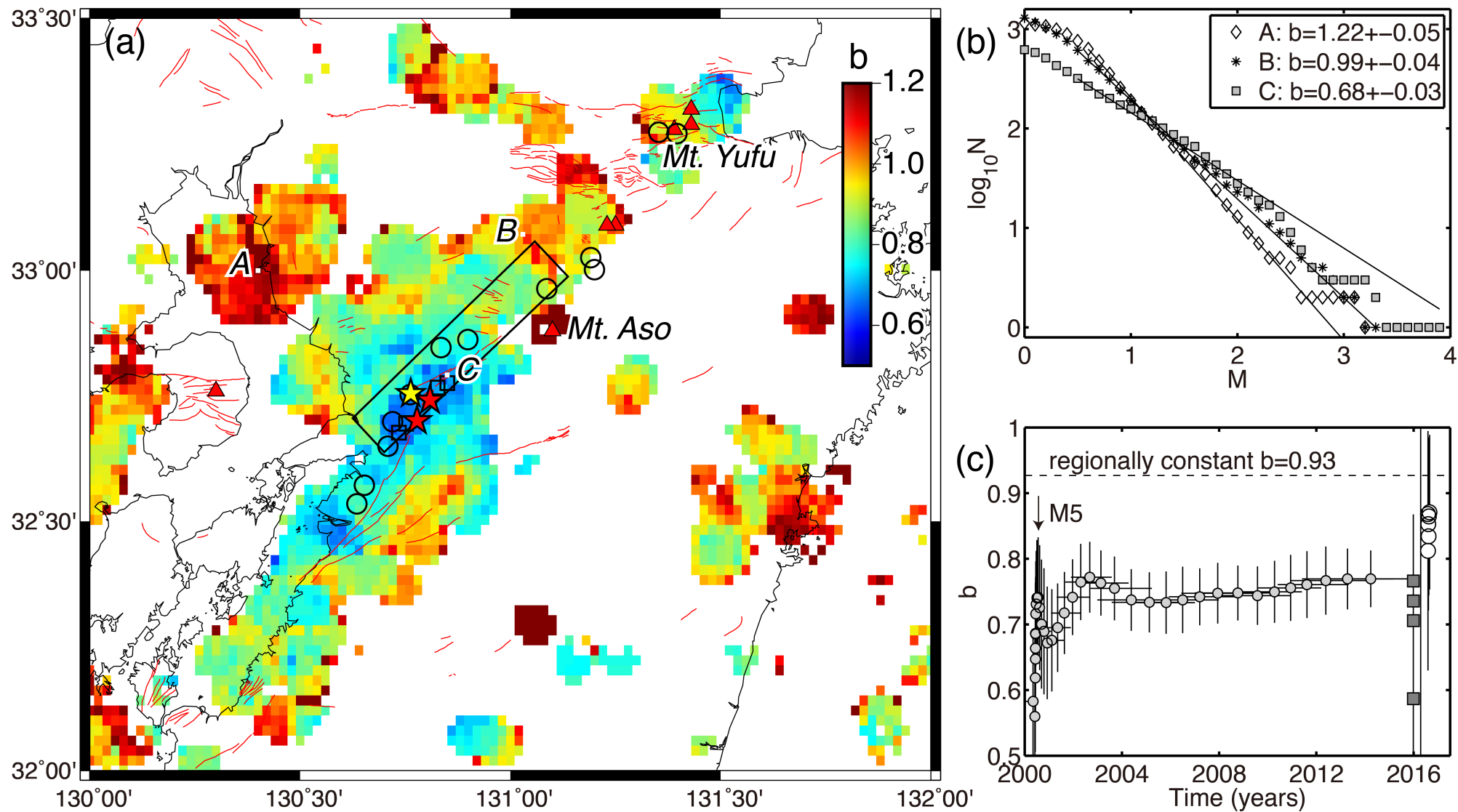

Fig. 1

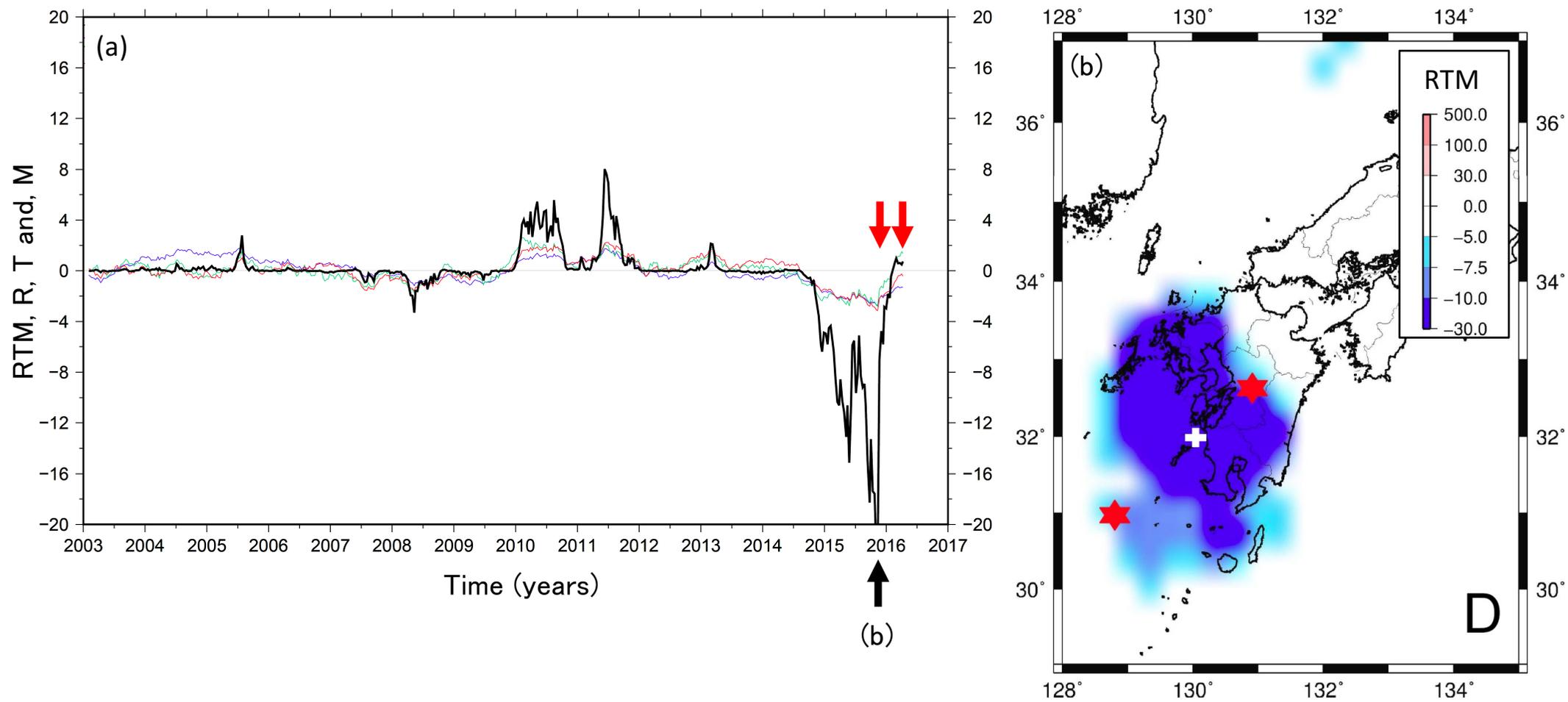

Fig. 2

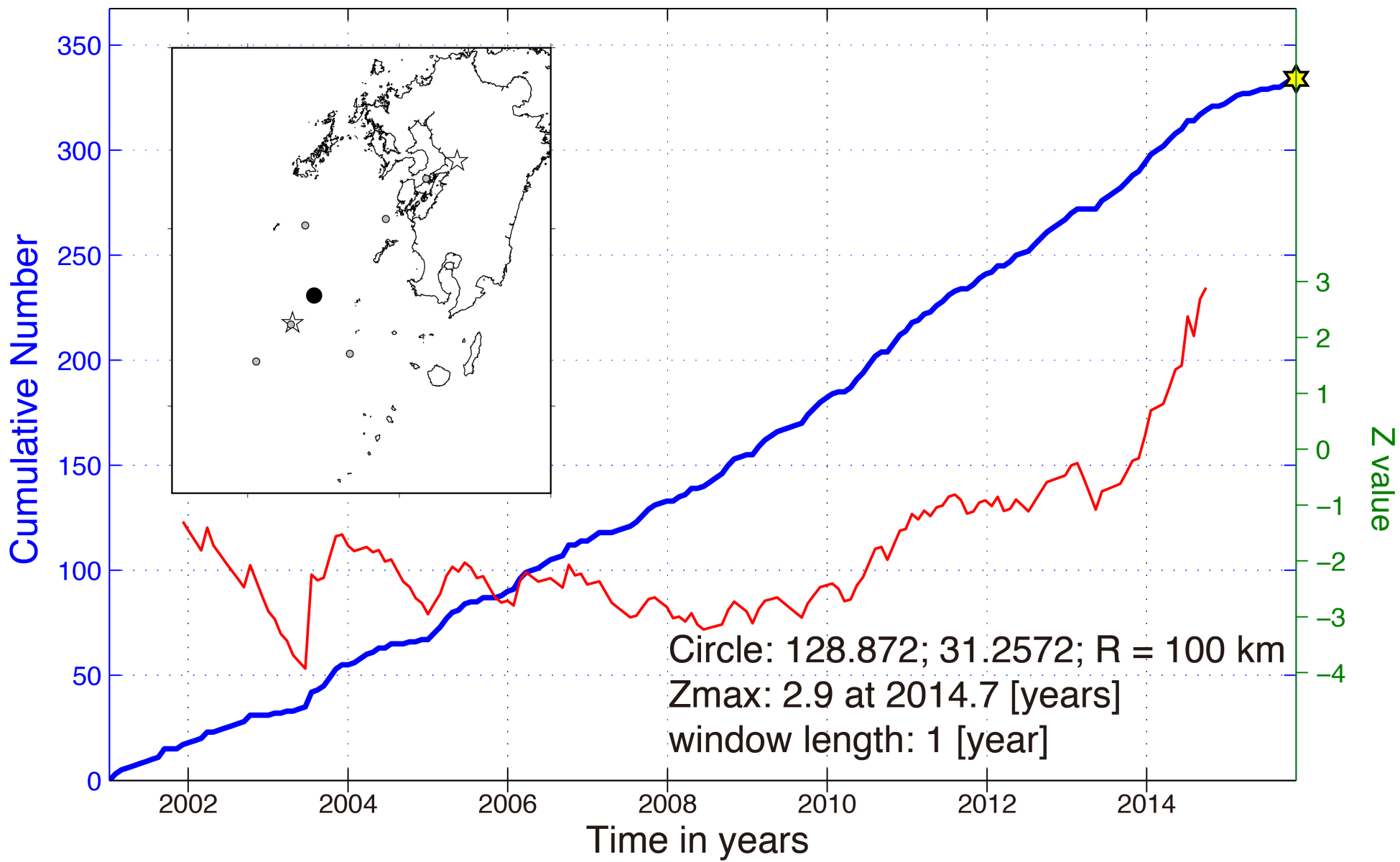

Fig. 3

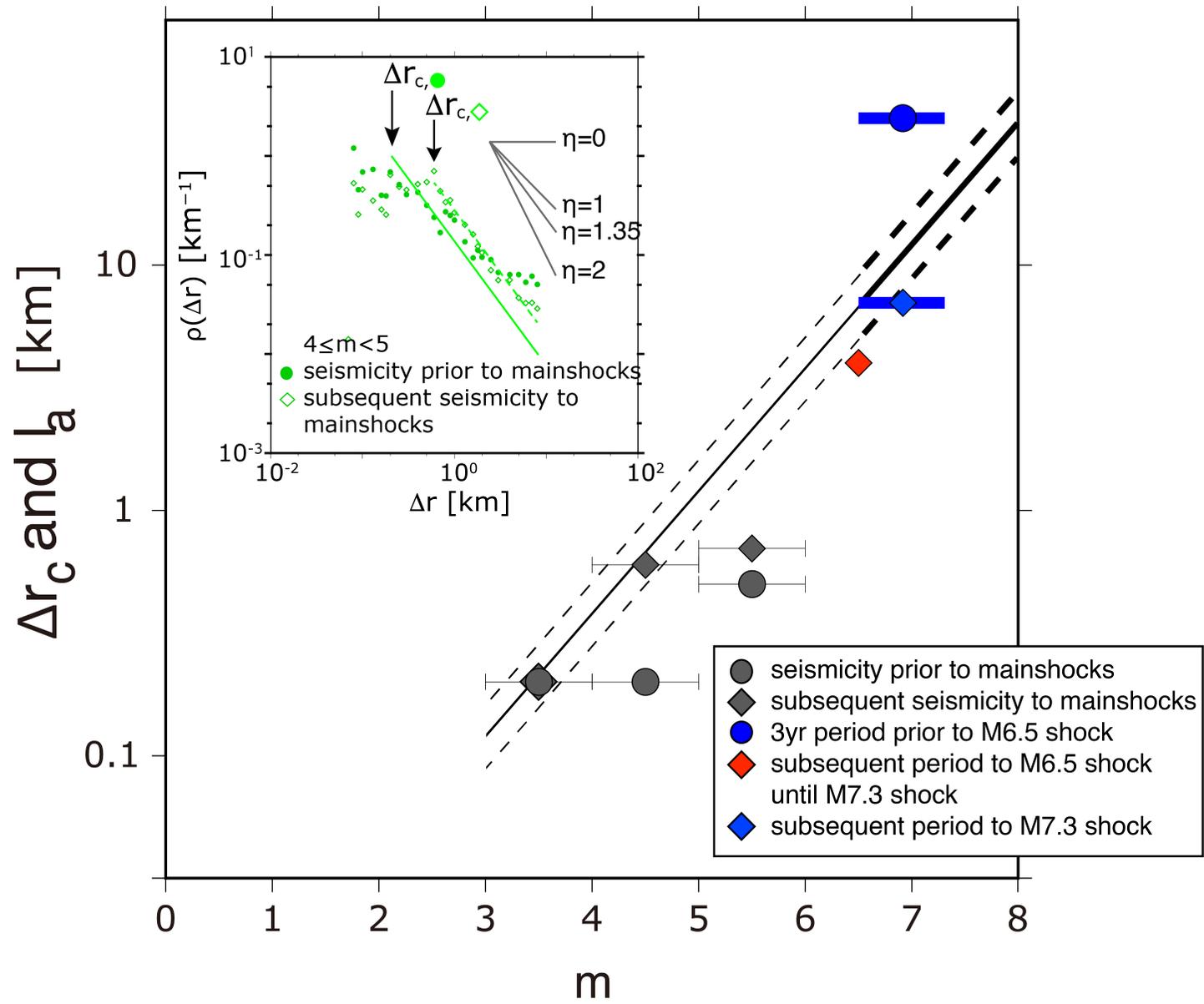

Fig. 4

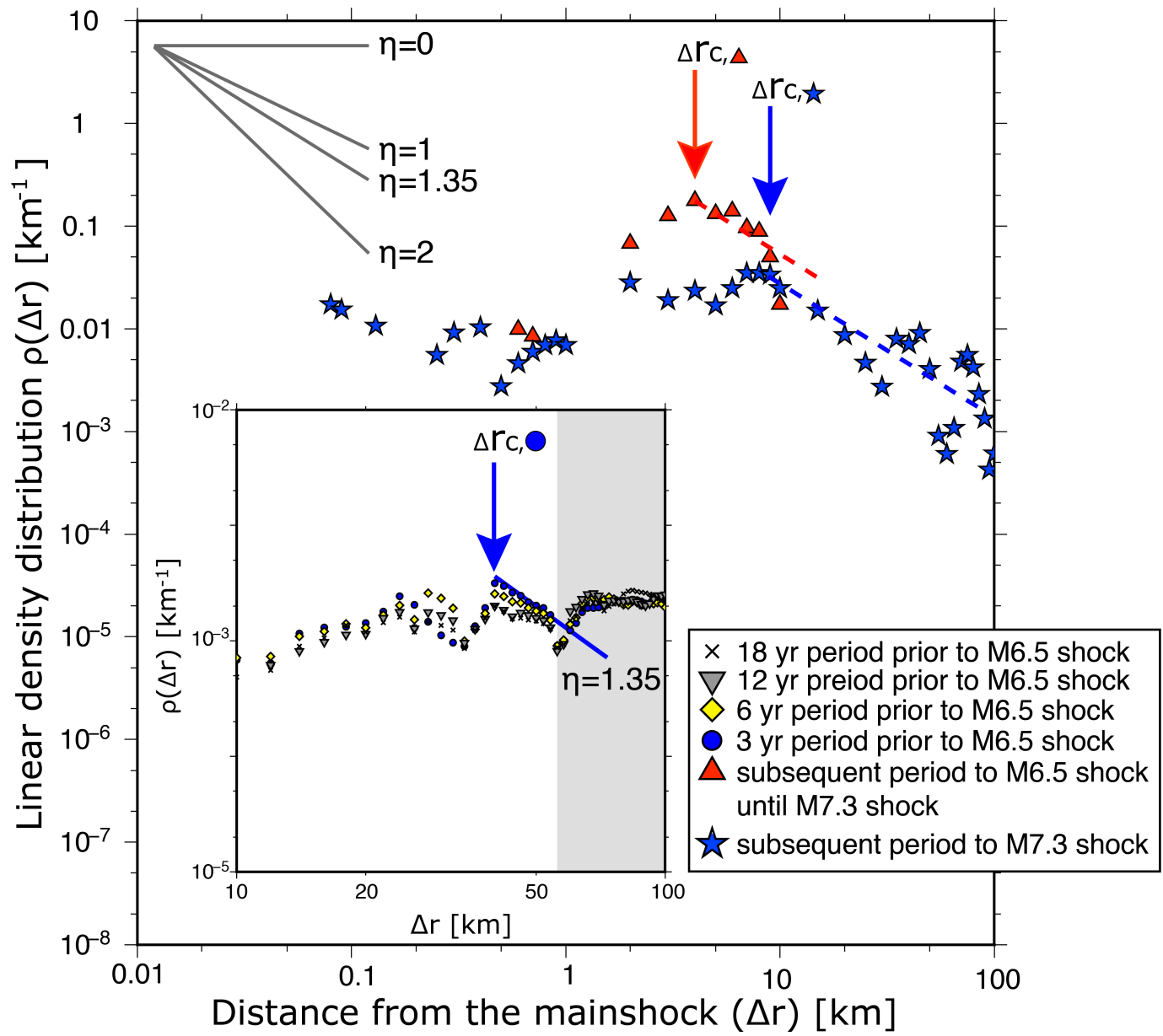

Fig. 5